\documentclass[a4paper,10pt,twoside]{cpc-hepnp}

\usepackage{multicol}
\usepackage{graphicx}
\usepackage{booktabs}
\usepackage{amssymb,bm,mathrsfs,bbm,amscd}
\usepackage[tbtags]{amsmath}
\usepackage{lastpage}
\usepackage{multirow}
\usepackage{CJK}

\begin{document}

\fancyhead[c]{\small Chinese Physics C~~~Vol. XX, No. X (201X)
XXXXXX} \fancyfoot[C]{\small 010201-\thepage}

\footnotetext[0]{}

\title{A digital CDS technique and the performance testing\thanks{Partially supported by National Natural Science
Foundation of China (10978002) }}

\author{LIU Xiao-Yan$^{1,2}$
\quad LU Jing-Bin$^{1}$
\quad YANG Yan-Ji$^{1,2}$
\quad LU Bo$^{2}$\\
\quad WANG Yu-Sa$^{2}$
\quad XU Yu-Peng$^{2}$
\quad CUI Wei-Wei$^{2}$
\quad LI Wei$^{2}$\\
\quad LI Mao-Shun$^{2}$
\quad WANG Juan$^{2}$
\quad HAN Da-Wei$^{2}$
\quad CHEN Tian-Xiang$^{2}$\\
\quad HUO Jia$^{2}$
\quad HU Wei$^{2}$
\quad ZHANG Yi$^{2}$
\quad ZHU Yue$^{2}$\\
\quad ZHANG Zi-Liang$^{2}$
\quad YIN Guo-He$^{2}$
\quad Wang Yu$^{2}$
\quad Zhao Zhong-Yi$^{2,3}$\\
\quad Fu Yan-Hong$^{2,3}$
\quad Zhang Ya$^{2,3}$
\quad MA Ke-Yan$^{1}$
\quad CHEN Yong$^{2;1)}$\email{ychen@mail.ihep.ac.cn}
}
\maketitle

\address{%
$^1$ College of Physics, Jilin University, No.2699, Qianjin Road, Changchun 130023, China\\
$^2$ Key Laboratory for Particle Astrophysics, Institute of High Energy Physics, Chinese Academy of Sciences (CAS), 19B Yuquan
Road, Beijing 100049, China\\
$^3$ School of Physical Science and Technology, Yunnan University, Cuihu North Road 2, Kunming, 650091, China\\
}

\begin{abstract}
Readout noise is a critical parameter for characterizing the performance of charge-coupled devices (CCDs), which can be greatly reduced by the correlated double sampling (CDS) circuit. However, conventional CDS circuit inevitably introduces new noises since it consists of several active analog components such as operational amplifiers. This paper proposes a digital CDS circuit technique, which transforms the pre-amplified CCD signal into a train of digital presentations by a high-speed data acquisition card directly without the noisy CDS circuit first, then implement the digital CDS algorithm through numerical method. The readout noise of 3.3 e$^{-}$ and the energy resolution of 121 eV@5.9keV can be achieved via the digital CDS technique.
\end{abstract}

\begin{keyword}
charge-coupled devices, readout noise, correlated double sampling
\end{keyword}

\begin{pacs}
29.30.Kv, 29.40.Wk, 29.85.-c
\end{pacs}

\footnotetext[0]{\hspace*{-3mm}\raisebox{0.3ex}{$\scriptstyle\copyright$}2013
Chinese Physical Society and the Institute of High Energy Physics
of the Chinese Academy of Sciences and the Institute
of Modern Physics of the Chinese Academy of Sciences and IOP Publishing Ltd}%

\begin{multicols}{2}

\section{Introduction}
Owing to its advantages on smaller size, lower power dissipation, wider response spectrum range, lower noise and higher resolution, Charge-coupled devices (CCDs) have been widely applied in industrial inspection, night vision, visible imaging, soft X-ray astronomical observations and so on. As a critical parameter that has to be considered in designing and operating CCD, the noises of CCD mainly come from two mechanisms, one comes from CCD itself, including shot noise, dark current noise and the transfer noise\cite{lab1}, and the other comes from the operation of CCD, such as output amplifier noise and the reset noise.

It is common to suppress the reset noise with the correlated double sampling (CDS) circuit. However, traditional CDS introduces new noises, and leaves some useful information behind, e.g. the original waveform. The digital CDS, e.g. Gach J L, 2003\cite{lab2}, is able to store all the initial information, which provides the possibility for varies of the backend data process. In practice, it shows a perfect performance on the reduction of the readout noise.

In this paper, a digital correlated double sampling circuit technique is proposed. A PCI-9846H data acquisition card, which is manufactured by ADLINK Inc., has been used to convert the pre-amplified CCD signal into the digital representations. The digital CDS system can record a large amount of data and the original waveform could be derived from these data for further analyzing of the signal characteristics. So the data processing could be optimized to get better performance such as a lower readout noise. The relationship between the readout noise of the digital CDS system and the sample number has been investigated. The results of the measured data show that the proposed digital CDS system is better than the conventional analog CDS system, which has been assembled on the low energy X-ray telescope (LE) of the Hard X-ray Modulation Telescope (HXMT)\cite{lab3}, whatever the CCD operating temperature is below $-110\,^{\circ}\mathrm{C}$, or the CCD is degraded after irradiation by proton.

\section{Reset noise and Correlated Double Sampling}
Charge generation, collection, transfer and measurement, make all of the operation of CCD. Taking two-phase CCD for example, the electronic schematic of CCD output region is illustrated on the left part in Fig.~\ref{fig1}, the right part in Fig.~\ref{fig1} is the simple diagram of output signal. Reset noise is generated by the periodic reset of the sense-node capacitance ($C_\mathrm{n}$) by the reset switch. Reset noise, $n_\mathrm{R}(e^{-})$, can be expressed as
\begin{eqnarray}
\label{eq2}
n_\mathrm{R}(e^{-}) &=& \frac{\sqrt{kTC}}{q},
\end{eqnarray}
where $k$ is the Boltzmann's constant ($1.38\times10^{-23}$ J/K), $T$ is the absolute temperature (K), $C$ is the capacitance (F), and $q$ is the electronic charge. Reset noise can also be called ¡°kTC noise¡±\cite{lab4}.
The noise voltage changes very quickly when reset switch is on, leaving an undefined level on the $C_\mathrm{n}$ after the switch is off. After the reset switch has turned off, the noise voltage does not change significantly over a pixel period. Reset noise can be removed or reduced by using CDS circuit technique which subtracts the sample taken from the signal level from the sample taken from the reference level.
However, conventional CDS circuit inevitably introduces new noises since it consists of several active analog components such as operational amplifiers. In this paper, we use digital CDS system instead of the conventional analog CDS.
\begin{center}
\includegraphics[origin=c,width=8cm]{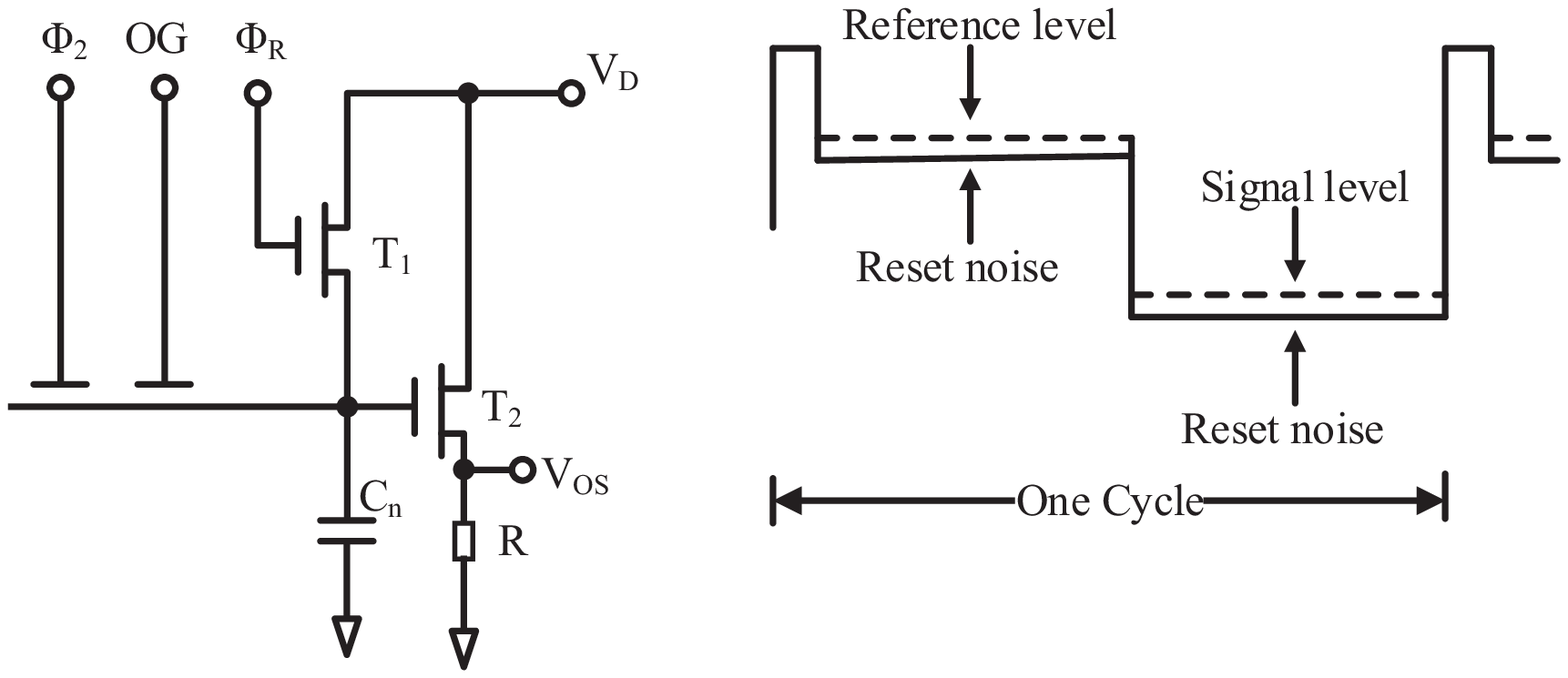}
\figcaption{\label{fig1} The schematic of CCD output region and output signal.}
\end{center}

\section{Brief introduction to the experiment equipment}
\subsection{The target CCD detector}
The CCD detector discussed in this paper is CCD236. It is a two-phased swept charge device (SCD) which is designed by the E2V Inc. for HXMT-LE, collaborated with the Institute of High Energy Physics (IHEP). CCD236 achieves a high transfer rate by abandoning the position information of the X-ray photons. Charges are first transferred to the electrodes on the diagonal along the way perpendicular of the electrodes, and then to the read-out section, which locates in the center of the chips, along the diagonal.
Besides the real output, there is a dummy output the same as the real one. As the dummy output only carries common-mode noise, the negative effect of common-mode noise can be greatly suppressed by subtracting that noise from the real output signal\cite{lab5}.
\subsection{The data acquisition card}
The PCI-9846H is a data acquisition card manufactured by the ADLINK inc., which is designed for digitizing high frequency and wide dynamic range signals, taking advantages in the rapid read-out speed and lower system noise. Its onboard 512 MB acquisition memory enables it to store the data of the waveform for a considerable long time. It has a high linearity 16-bit A/D converter, and has a sampling rate up to 40 million samples per second (MS/s). In addition,
 the analog voltage range is $\pm$5V\cite{lab6}.
\subsection{The readout system using the digital CDS}

Digital CDS system block diagram is shown as Fig.~\ref{fig2}, which is made up of three parts: pre-amplifier, voltage buffer and data acquisition system. The voltage gain of pre-amplifier is 11x. The voltage buffer ensures the accuracy of the signal amplitude. The PCI-9846H digitizes the analog input at a certain rate, and sends the conversion results to computer via PCI (peripheral component interconnection) for backend data processing.
\begin{center}
\includegraphics[origin=c,width=8cm]{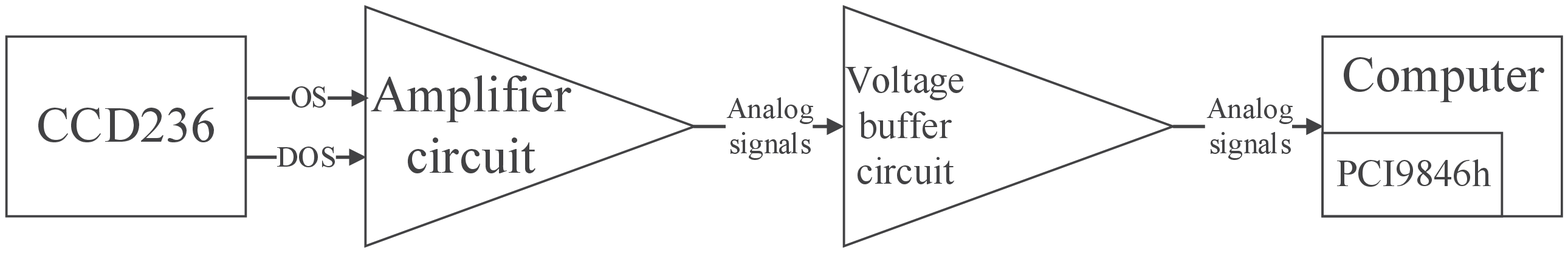}
\figcaption{\label{fig2} The digital CDS system block diagram.}
\end{center}

\section{The results of performance testing}
\subsection{Testing results at low CCD operating temperature}
To investigate the performance of the digital CDS system, a test has been carried out with the CCD operated below about $-110\,^{\circ}\mathrm{C}$, as the effect of the dark current from the CCD can be ignored at such a low temperature. The driving clock for the CCD is 83.3kHz, while the digital data acquisition system is sampling at 40 MS/s (million samples per second). An Fe-55 radioactive source is used to measure the energy resolution.

As shown in Fig.~\ref{fig3}, the waveform is reconstructed from total 480 sample points. Due to such non-idealities as clock feed-through, parasitic effects as well as signal integrity problems, glitches are prominent at the edges of driving clocks. However, only the two steady parts every cycle make sense, since signals in these two parts are completely settled. A certain number of sample points are selected as effective points for processing.
\begin{center}
\includegraphics[origin=c,width=8cm]{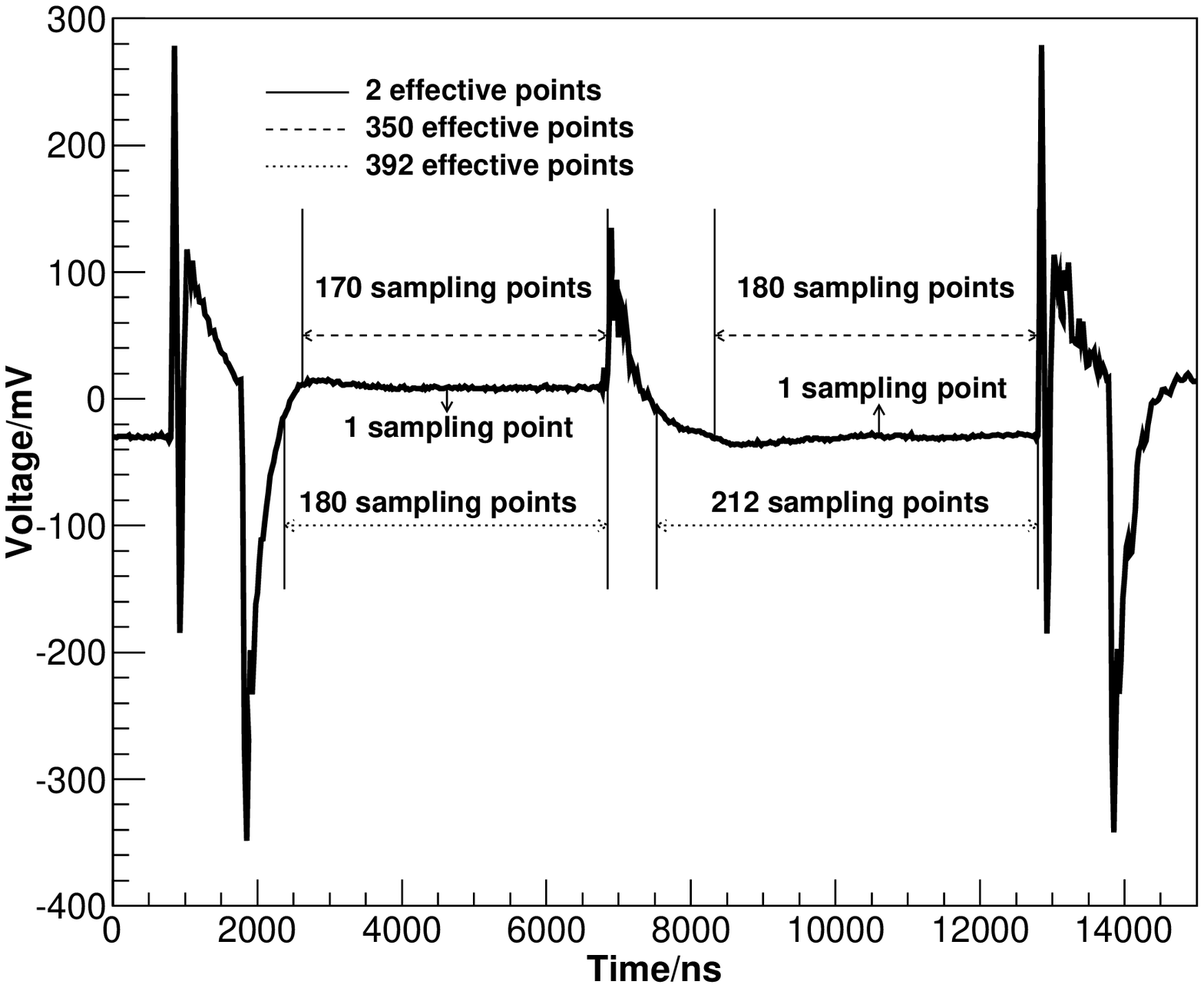}
\figcaption{\label{fig3} The typical signal waveform.}
\end{center}

The flat section of the curve in Fig.~\ref{fig3} is selected on the basis of that the difference of the voltage between two adjacent sampling points is no more than 40 mV. Taking the section of the signal level as an example, the initial point (the 389th point) is exactly placed in the middle of the flat section, the second one is located just left to the initial one (i.e. the 388th point), the third on is situated just right to the initial one (i.e. the 390th point), and so on. All the points right to the 477th point are ignored. The way to select points on the reference level is just the same, and the initial point in this case is the 150th point. During the following numerical calculation, the number of effective points increases from 2 to 392 in order to find out the optimal number. Eventually, the voltage amplitude of the incoming X-ray photons can be obtained by the equation
\begin{eqnarray}
\label{eq3}
V_\mathrm{ph} &=& \overline{V_\mathrm{Ref}} - \overline{V_\mathrm{Signal}},
\end{eqnarray}
where $\overline{V_\mathrm{Signal}}$ is the average voltage of the sampling points on the flat section of the signal level, and $\overline{V_\mathrm{Ref}}$ is the average voltage of the sampling points on the flat section of the reference level.

\begin{center}
\includegraphics[origin=c,width=8cm]{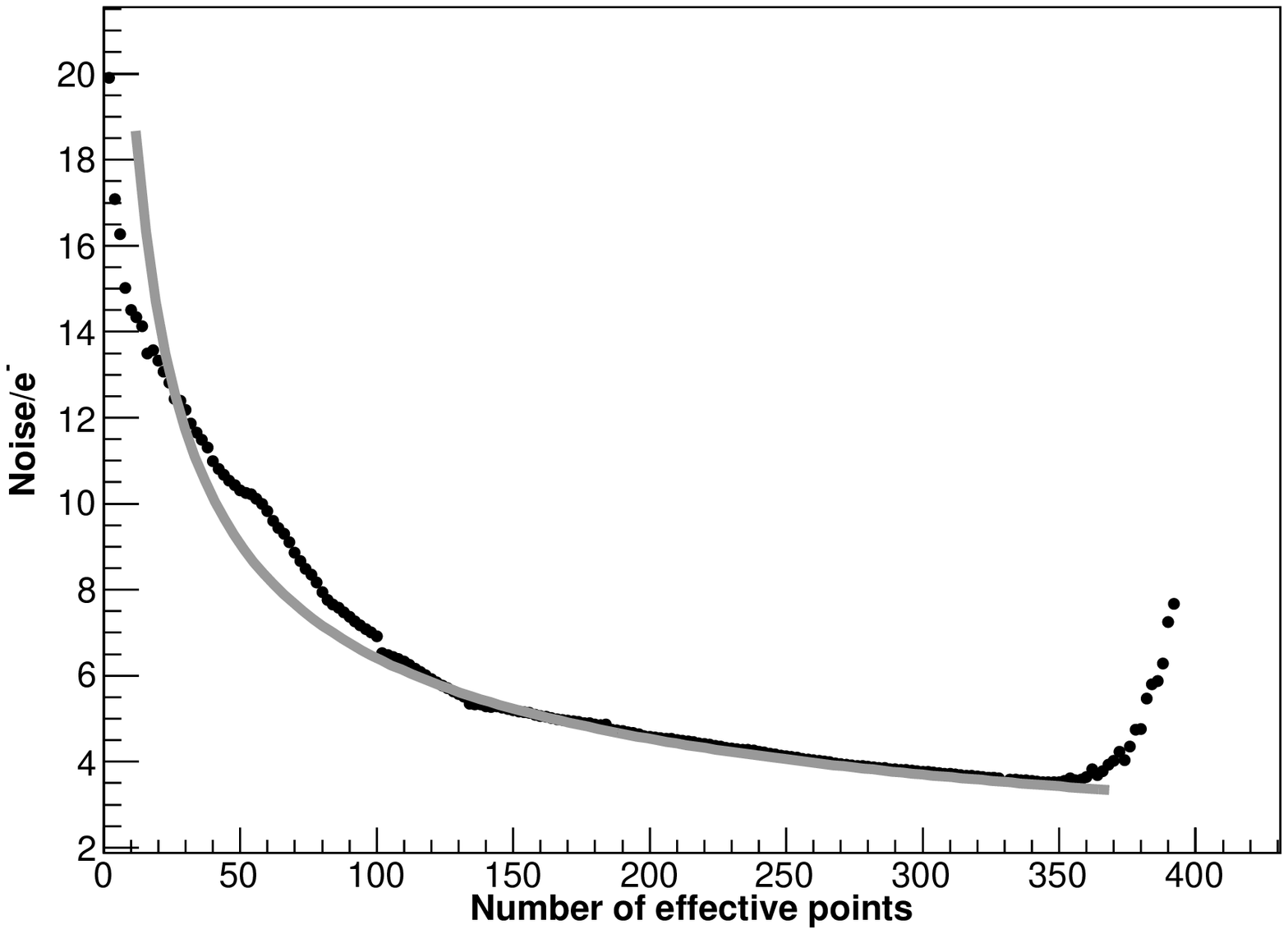}
\figcaption{\label{fig4} Readout noise vs. the number of effective sampling points under 83.3kHz operating frequency.}
\end{center}

As shown in Fig.~\ref{fig4}, below 350 samples, the readout noise decreases with the effective points increasing. The function (Eq.~\ref{eq3}) has been used to fit the relationship between the readout noise and the number of effective sampling points, which describes the error of the mean value for the independent sampling data.

We try to use Eq.~\ref{eq4} to fit the data in Fig.~\ref{fig4}.
\begin{eqnarray}
\label{eq4}
n &=& \frac{A_\mathrm{0}}{\sqrt{N}},
\end{eqnarray}
where $n$ is readout noise ($\mathrm{e^{-}}$), and $N$ is the number of effective sampling points. The fitting curve is shown in Fig.~\ref{fig4} as the solid line, with the value of $A_\mathrm{0}$ $64.1 \pm0.6 \mathrm{e^{-}}$. When only one point are sampled on each level ($N=2$), the noise is about $20 \mathrm{e^{-}}$, which is significantly lower than the value of the reset noise (53$\mathrm{e^{-}}$) calculated by Eq.~\ref{eq2}, where $C$ is 0.032 pF for CCD236 and $T$ is 160 K. However, the declination of the data in Fig.~\ref{fig4} indicates that there are also other sources of the noises or interferences besides the reset noise.

Eq.~\ref{eq4} cannot fit the data well at both ends. In detail, the readout noise increases again when the number of effective points exceeds 350, because the data beyond the steady parts are used. The ranges of 350, 392 effective points are shown in Fig.~\ref{fig3}. Obviously, the optimal effective point number is around 350, so we select this number in all the test below in this paper. As the sampling points are fewer, shown in left-hand side of Fig.~\ref{fig4}, the declining slope of the data is significantly shallower than that of the $1/\sqrt{N}$. It is suggested that the sampling data are not independent in this situation.

We calculated the cross-correlation function of the voltage of each sampling point and the initial sampling point (i.e. the 389th point). As shown in Fig.~\ref{fig5}, the sampling points located faraway are weakly correlated, while the adjacent sampling points show sharp oscillation, from strongly correlated to anti-correlated. It indicates that the readout noise is probably not the white noise.

\begin{center}
\includegraphics[origin=c,width=8cm]{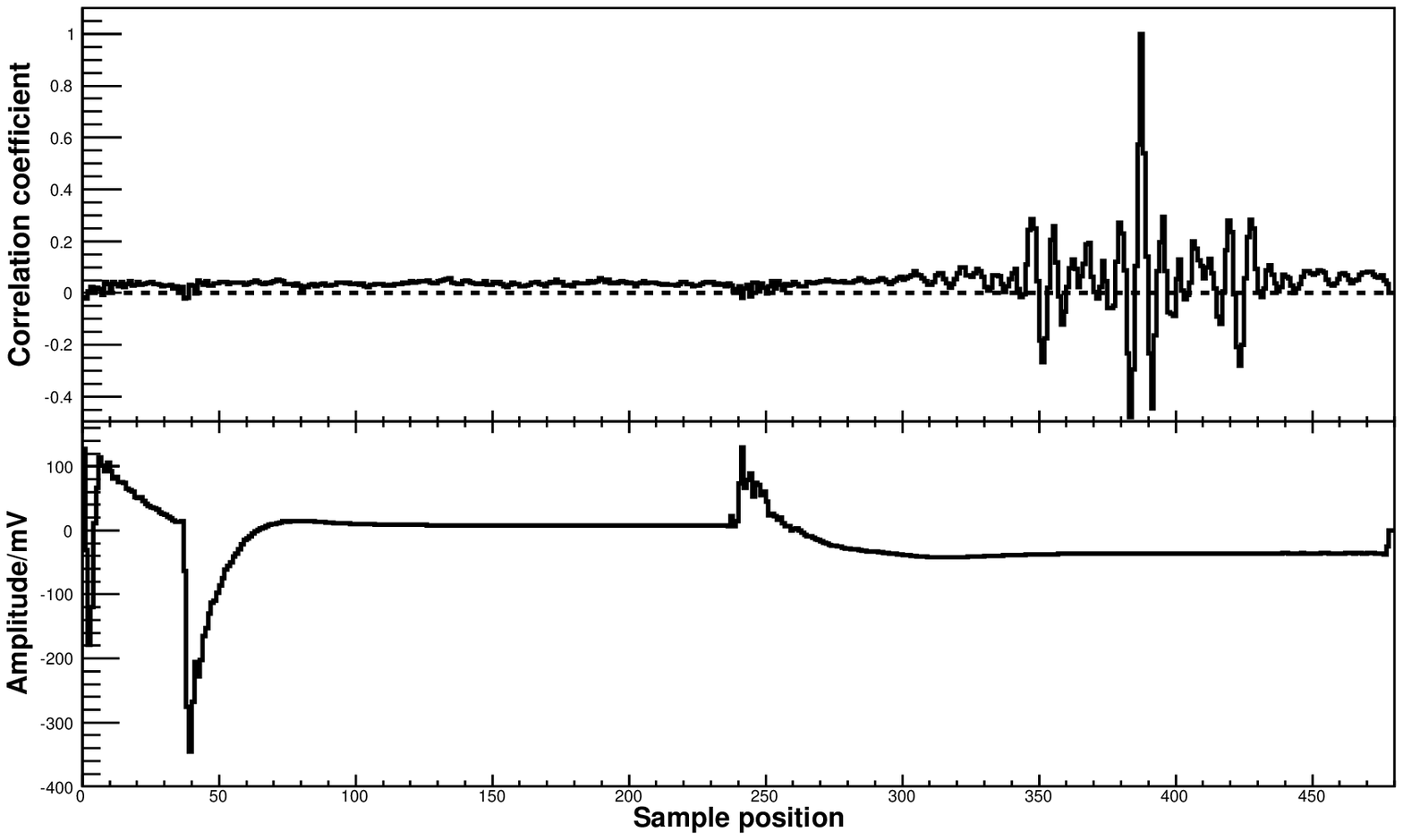}
\figcaption{\label{fig5} The cross-correlation between the voltages of each sampling point and the initial sampling point (the 389th point). The upper plot is the cross-correlation data, and the lower plot is the mean value of the waveform for reference.}
\end{center}

For this purpose, the power spectrum of the readout noise (the part of the signal level) are calculated by the fast Fourier transform (FFT) method. In the power spectrum (Fig.~\ref{fig6}), there are low-frequency noises substantially, most likely $1/f^{\mathrm{\alpha}}$ noise with $\alpha$ from 1 to 2. The lower frequency noise component may be responsible to the weak correlation in long distance. On the other hand, it is found that there is an enhancement at about 5 MHz in the power spectrum, which can explain the strong correlation in short distance. The lower frequency noise may come from the onchip field effect transistor (FET) of the CCD readout circuit\cite{lab7}, and the $\mathrm{\sim}$5 MHz noise is probably caused by the outside interference or the CCD itself.

\begin{center}
\includegraphics[origin=c,width=8cm]{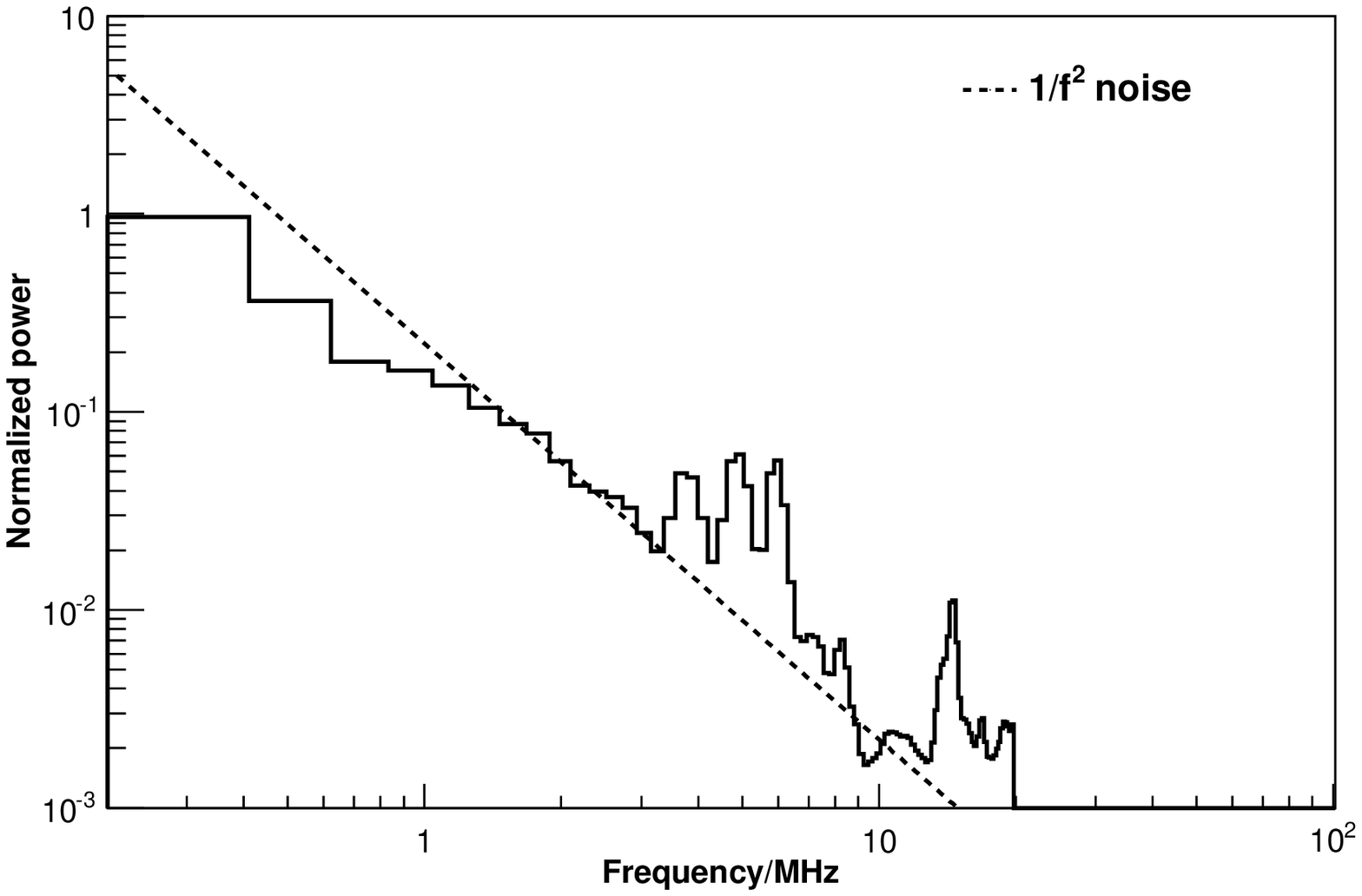}
\figcaption{\label{fig6} The normalized power spectrum of the readout noise in the part of the signal level, with a $1/f^{2}$ noise shown in dashed line for reference.}
\end{center}

To determine the performance, especially the energy linearity, an X-ray tube with the copper target is used. From the spectrum which is shown in Fig.~\ref{fig7}, four distinct peaks besides the noise peak are found. The integral nonlinearity (INL) is satisfied with the value of $0.4\%$.
\begin{center}
\includegraphics[origin=c,width=8cm]{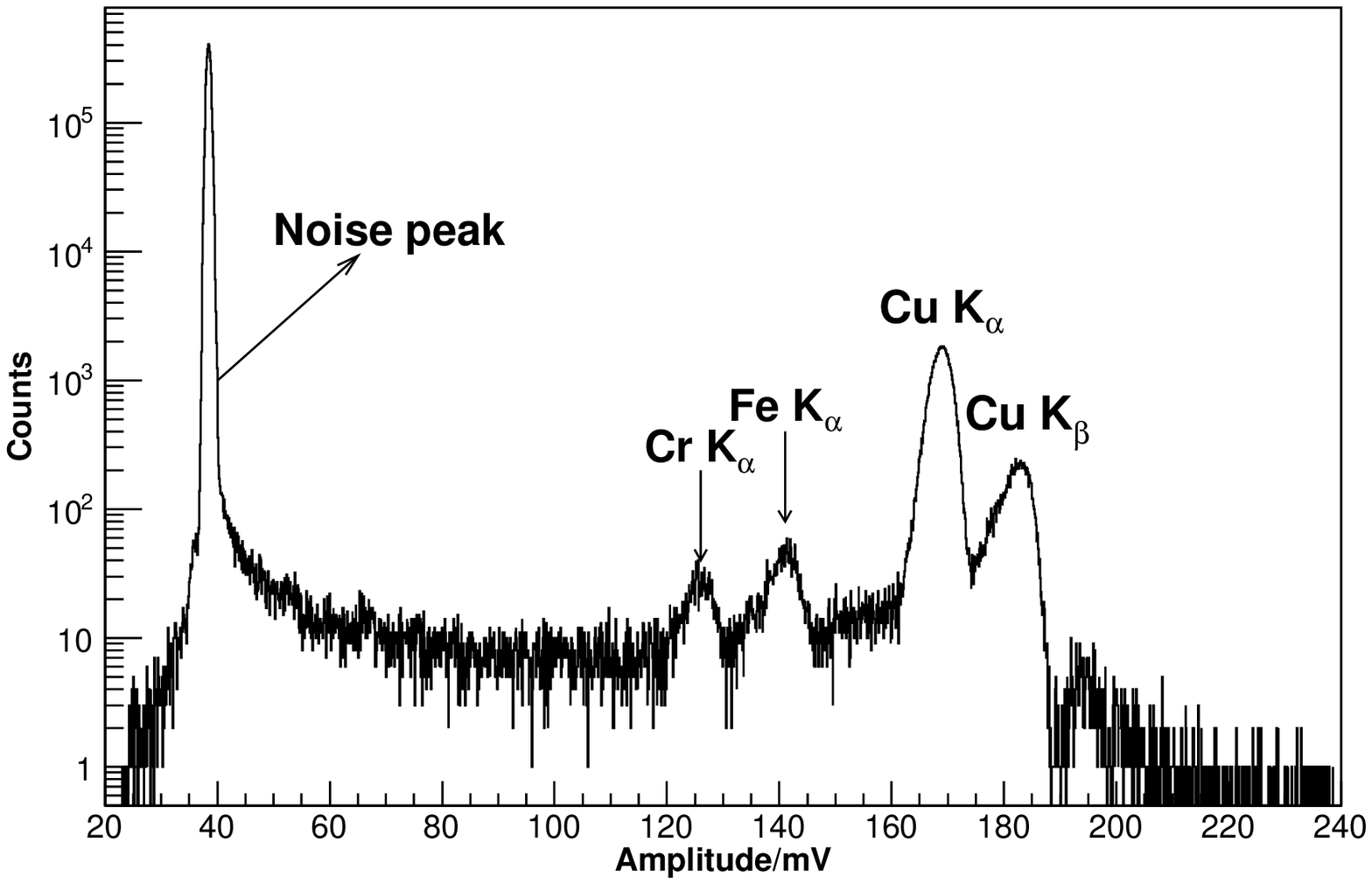}
\figcaption{\label{fig7} The spectrum from a copper target bombarded with X-rays generated by an X-ray tube.}
\end{center}
\begin{center}
\includegraphics[origin=c,width=8cm]{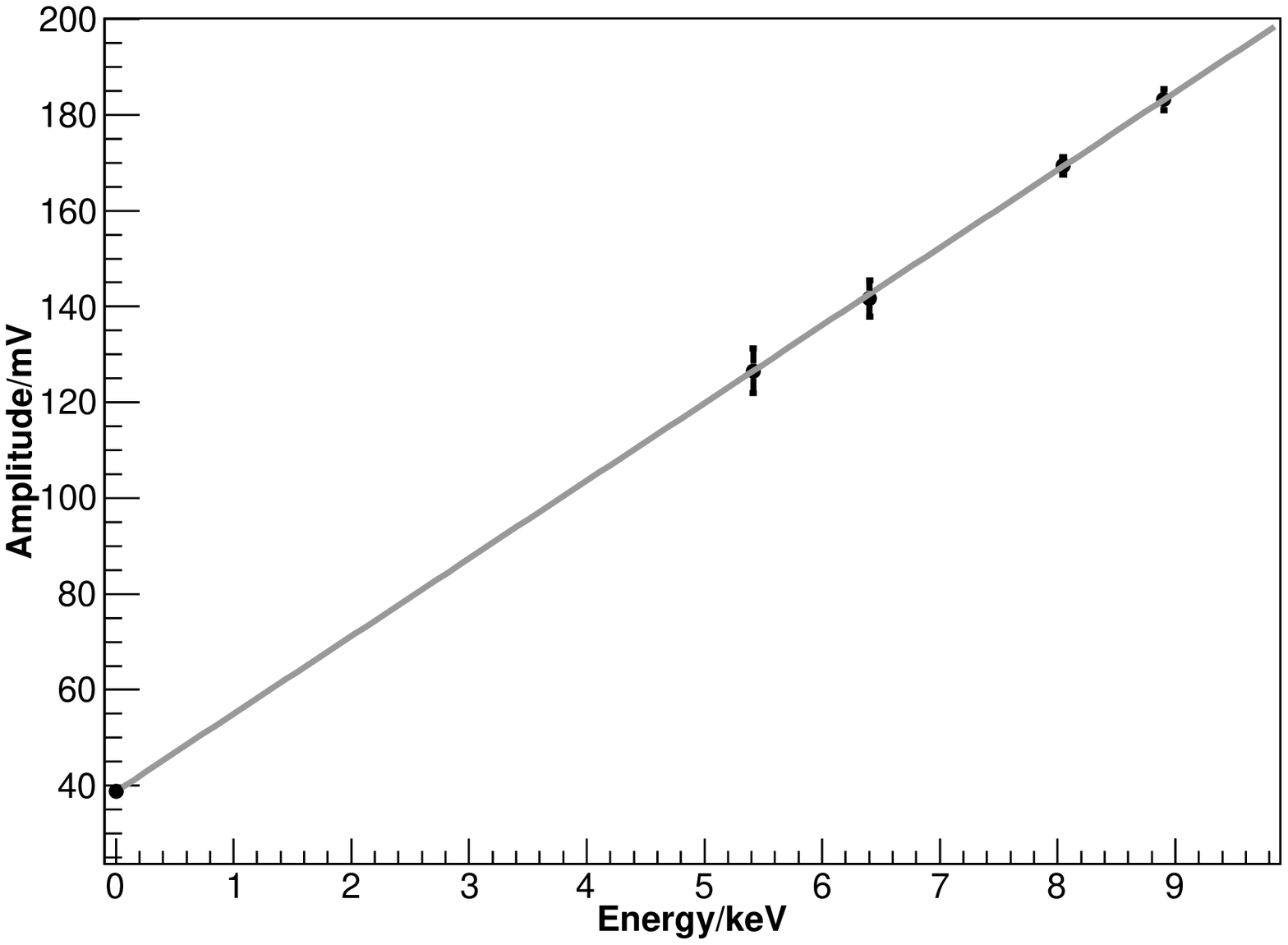}
\figcaption{\label{fig8} The integral nonlinearity of the digital CDS system.}
\end{center}

To investigate the energy resolution of the system, the test has been conducted with an Fe-55 radioactive source. The spectrum is shown in Fig.~\ref{fig9}. Fitting the noise peak and Mn K$_{\alpha}$ (5.9 keV) peak by Gaussian, the parameters as the mean value and the standard deviation can be obtained, the energy resolution of the CCD detector system is calculated to be 121.0$\pm$0.4eV@5.9keV (FWHM), and the readout noise is 3.342$\pm$0.002e$^{\mathrm{-}}$.
\begin{center}
\includegraphics[origin=c,width=6cm]{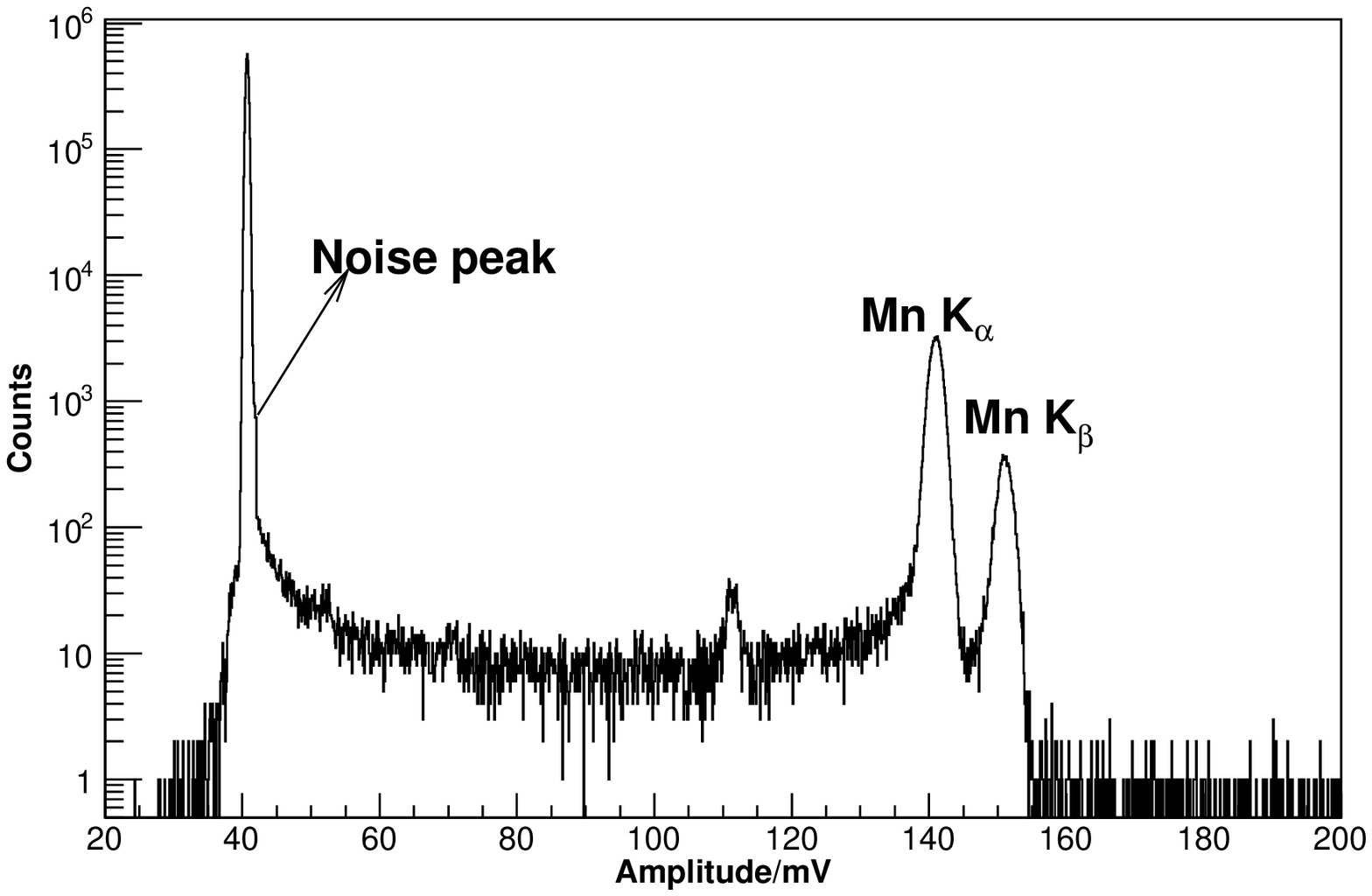}
\figcaption{\label{fig9} An Fe-55 spectrum of CCD236 at $-113\,^{\circ}\mathrm{C}$.}
\end{center}

Within the temperature range of $-116\,^{\circ}\mathrm{C}$ to $-110\,^{\circ}\mathrm{C}$, the readout noise and energy resolution of the digital CDS system and the conventional CDS readout system under the same condition are compared in Table~\ref{tab1}, which illustrates that the performance of digital CDS system is better than that of the conventional CDS system.
\end{multicols}
\begin{center}
\tabcaption{\label{tab1} The performance comparison between two systems.}
\footnotesize
\begin{tabular}{c|c|c|c|c}
\toprule
\multirow{3}{*}{CCD operational} & \multicolumn{2}{|c|}{Energy resolution(FWHM)/eV} & \multicolumn{2}{c}{Readout noise/e$^{-}$}\\
& Digital CDS& Conventional CDS & Digital CDS & Conventional CDS\\
temperature $/\,^{\circ}\mathrm{C}$ & system & system & system & system \\
\hline
-110.3 & 123.4$\pm$0.4 &131.9$\pm$1.1 &3.487$\pm$0.002 & 5.251$\pm$0.002\\
-113.0 & 121.0$\pm$0.4 &131.0$\pm$1.1 &3.342$\pm$0.002 & 5.428$\pm$0.002\\
-115.1 & 121.7$\pm$0.4 &131.4$\pm$1.1 &3.331$\pm$0.002 & 5.208$\pm$0.002\\
-116.2 & 121.3$\pm$0.4 &129.7$\pm$1.0 &3.329$\pm$0.002 & 5.416$\pm$0.002\\
\bottomrule
\end{tabular}
\end{center}
%\ruledown \vspace{0.5cm}
\begin{multicols}{2}
\subsection{The performance comparison between two systems after proton irradiation}
The readout noise and energy resolution of a proton-irradiated CCD236, with a dose of $3\times10^{8} \mathrm{p/cm}^{2}$, are measured with the temperature ranging from $-60\,^{\circ}\mathrm{C}$ to $-10\,^{\circ}\mathrm{C}$\cite{lab8}.

As shown in Fig.~\ref{fig11}, the readout noise of the digital CDS system is about 1 $\mathrm{e^{-}}$ lower than that of the conventional CDS readout system in the temperature range from $-60\,^{\circ}\mathrm{C}$ to $-30\,^{\circ}\mathrm{C}$. And the difference of readout noise gradually increases to 4 $\mathrm{e^{-}}$ when the temperature warms up to -10 ¡æ. In addition, the energy resolution of the digital CDS system also shows a better result similarly (Fig.~\ref{fig12}). The performance of digital CDS system is better than conventional CDS system after proton irradiation. Therefore, it is a promising way to suppress the negative effects of the proton-irradiation damage to the CCDs.
\begin{center}
\includegraphics[origin=c,width=8cm]{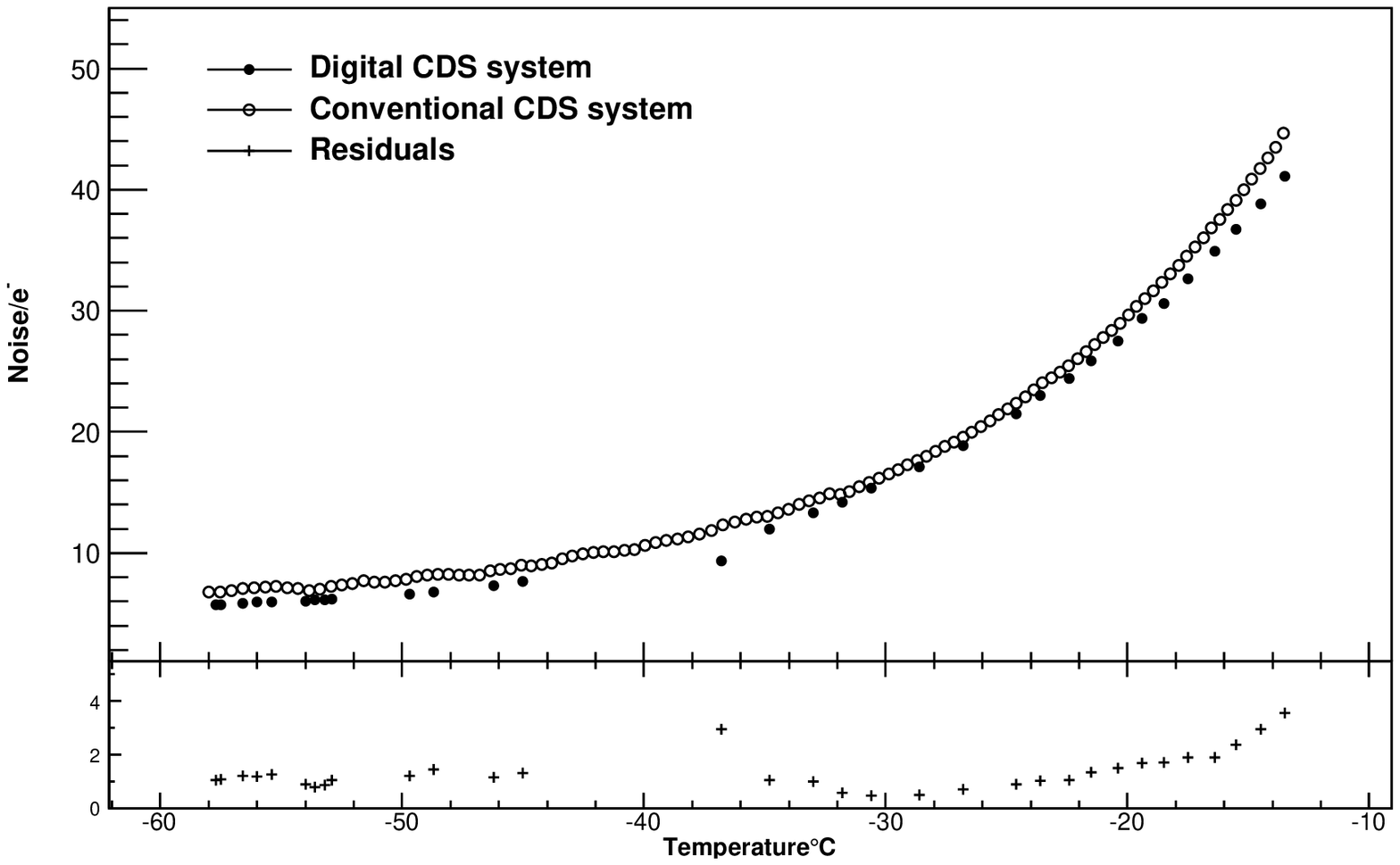}
\figcaption{\label{fig11} The comparison of the readout noise. The open circles, the solid circles and the plus signs (`+') represent the data of the digital CDS system, the conventional CDS system and the residuals, respectively.}
\end{center}
\begin{center}
\includegraphics[origin=c,width=8cm]{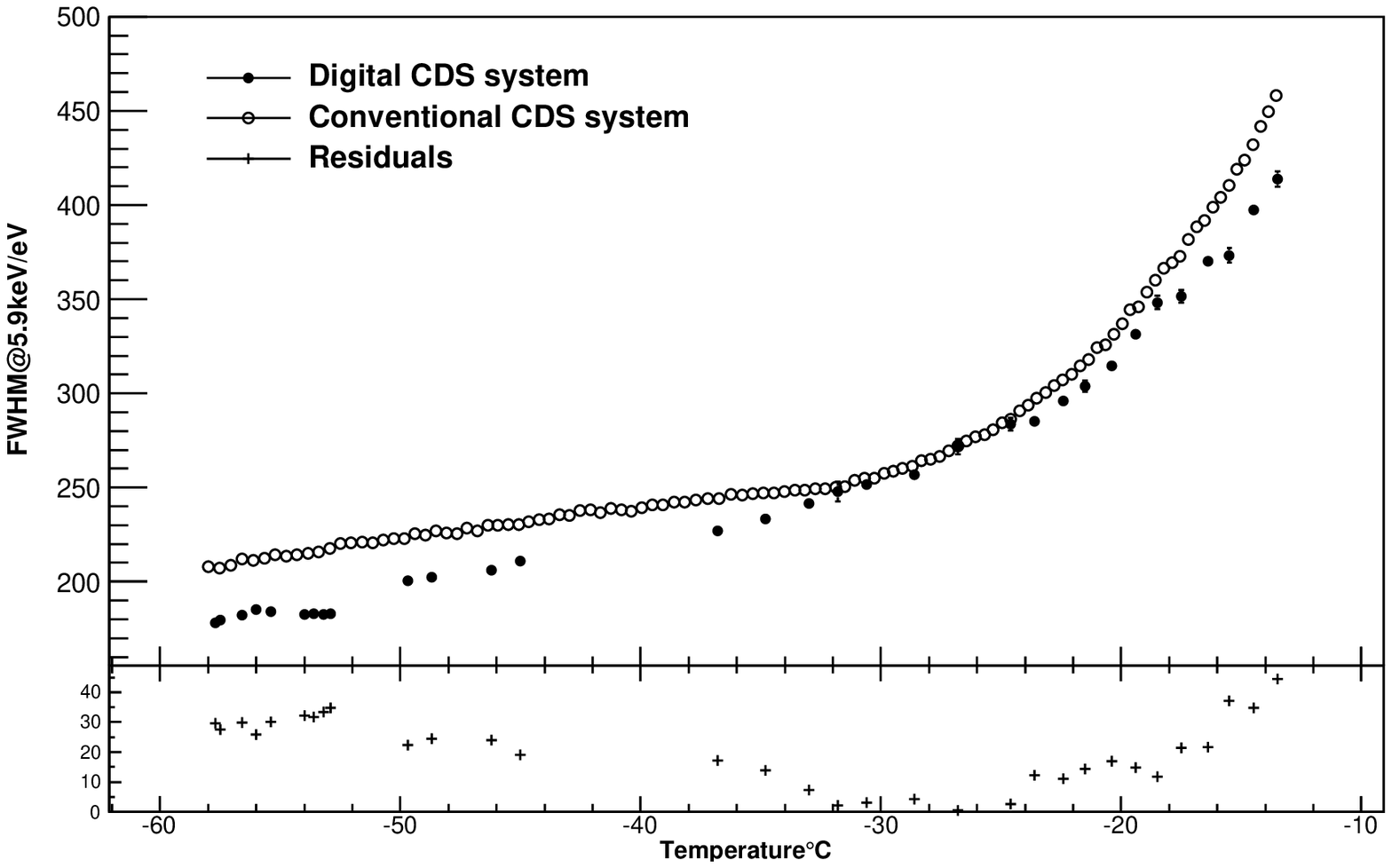}
\figcaption{\label{fig12} The same as Fig.~\ref{fig11}, but the comparison of the energy resolution.}
\end{center}
\section{Conclusions}
In this paper, we have proposed a digital CDS system. Under the same condition, the readout noise is reduced monotonously as the number of sampling points per cycle increases to 350. And from the waveform obtained from the digital CDS system, we have analysed the sources of the noise by the power spectrum. Compared to the conventional CDS system, the digital CDS system has a lower readout noise and a better energy resolution, not only under very low CCD operating temperatures but also on the CCD after the proton-irradiation. Additionally, the digital CDS system shows a satisfied result of INL. In summary, the digital CDS is a promising technique to reduce the readout noise of CCDs, to suppress the effect of the interferences, and to investigate the properties of the noise.
\\

\end{multicols}

\clearpage

\end{document}